\documentclass[aps,prl,reprint, superscriptaddress]{revtex4-2} % preprint
\usepackage{amsfonts}
\usepackage{amssymb}
\usepackage{amsthm,amsmath}
\usepackage{mathrsfs}
\usepackage{indentfirst}
\usepackage{multirow}
\usepackage{graphicx}
\usepackage{amsmath}
\usepackage[amssymb]{SIunits}
\usepackage{amssymb}
\usepackage{color}
\usepackage[usenames,dvipsnames]{xcolor}
\usepackage{ulem}
\usepackage{cancel}
\usepackage{natbib}
\usepackage[T1]{fontenc}
\usepackage[colorlinks=false, bookmarks=true]{hyperref}
\usepackage{bm}
\usepackage{booktabs}
\usepackage{multirow}
\begin{document}

\title{How roughness and thermal properties of a solid substrate determine the Leidenfrost temperature: Experiments and a model}

\author{Yuki Wakata}
\affiliation{Center for Combustion Energy, Key Laboratory for Thermal Science and Power Engineering of Ministry of Education, Department of Energy and Power Engineering, Tsinghua University, 100084 Beijing, China}
\author{Ning Zhu}
\affiliation{Center for Combustion Energy, Key Laboratory for Thermal Science and Power Engineering of Ministry of Education, Department of Energy and Power Engineering, Tsinghua University, 100084 Beijing, China}
\author{Xiaoliang Chen}
\affiliation{Center for Combustion Energy, Key Laboratory for Thermal Science and Power Engineering of Ministry of Education, Department of Energy and Power Engineering, Tsinghua University, 100084 Beijing, China}
\author{Sijia Lyu}
\affiliation{Center for Combustion Energy, Key Laboratory for Thermal Science and Power Engineering of Ministry of Education, Department of Energy and Power Engineering, Tsinghua University, 100084 Beijing, China}
\author{Detlef Lohse}
\email{d.lohse@utwente.nl}
\affiliation{Physics of Fluids Group, MESA$^{+}$ Institute  and J.M. Burgers Centre for Fluid Dynamics, University of Twente, P.O. Box 217, 7500AE Enschede, The Netherlands}
\affiliation{Max Planck Institute for Dynamics and Self-Organization, 37077 G\"ottingen, Germany}
\author{Xing Chao}
\email{chaox6@tsinghua.edu.cn}
\affiliation{Center for Combustion Energy, Key Laboratory for Thermal Science and Power Engineering of Ministry of Education, Department of Energy and Power Engineering, Tsinghua University, 100084 Beijing, China}
\author{Chao Sun}
\email{chaosun@tsinghua.edu.cn}
\affiliation{Center for Combustion Energy, Key Laboratory for Thermal Science and Power Engineering of Ministry of Education, Department of Energy and Power Engineering, Tsinghua University, 100084 Beijing, China}
\affiliation{Department of Engineering Mechanics, School of Aerospace Engineering, Tsinghua University, Beijing 100084, China}

\date{\today}

\begin{abstract} 
In this Letter, we systematically investigate the Leidenfrost temperature for hot solid substrates with various thermal diffusivities and surface roughnesses.
Based on the experimental results, we build a phenomenological model that considers the thermal diffusivity of a solid substrate and derive a relationship between the surface roughness and the resulting vapor film thickness. 
The generality of this model is supported by experimental data for different liquids and solid substrates. Our model thus allows for a theoretical prediction of the Leidenfrost temperature and develops a comprehensive understanding of the Leidenfrost effect.
\end{abstract}

\maketitle

Drops normally boil when deposited on surfaces with temperature above the boiling point of the liquid. However, when the surface temperature exceeds the Leidenfrost temperature $T_L$, a vapor layer generated by the evaporation levitates the drop, preventing contact and reducing the heat transfer between the drop and the surface \cite{Biance2003, Quere2013}. 
This effect, known as the Leidenfrost effect, has received much attention in recent years, both for its beauty and the rich physics involved in it \cite{Chantelot2021, Bouillant2021Self, Sijia2021On} and for its various practical applications \cite{Kim2007, Liang2017, Vakarelski2016, Gauthier2019, Bouillant2018, Abdelaziz2013}.
Therefore, learning and regulating the transition temperature $T_L$ is of scientific and practical value.

It has been observed that the Leidenfrost temperature $T_L$ of a static liquid drop on a flat surface is affected by several factors, including the properties of the solid surface (thermal properties, surface roughness, surface structure   \cite{Nagai1996,Kim2011,Jiang2022Inhibiting,Wu2021}), the ambient condition (pressure and temperature \cite{Celestini2013,VanLimbeek2021}) and the properties of the liquid \cite{Baumeister1973}, including the contact angle on the substrate \cite{Bourrianne2019,Panchanathan2021}.
While theoretical models have been developed for $T_L$ in terms of film stability \cite{Bernardin1999, Panzarella2000, Aursand2018, Zhao2020, Cai2020a}, the common approaches often assume that the substrate is isothermal or of uniform temperature, which is at odds with the finding that surfaces can be cooled by the floating Leidenfrost drop, as quantitatively shown in Ref. \cite{VanLimbeek2017}. However, the influence of the thermal properties of the solid on $T_L$ has not yet been systematically studied. 

In this Letter, we show how the thermal properties and surface roughness of a solid substrate affect the Leidenfrost effect and the Leidenfrost temperature $T_L$. 
The surface cooling caused by a Leidenfrost drop is shown with the help of infrared imaging, and Leidenfrost temperatures of water and ethanol drops on various surfaces are obtained with the aim to understand the dependence on the solid thermal diffusivity $\alpha$ and the surface roughness $S_\mathrm{a}$. 
A theoretical model describing the surface cooling effect is established and the influence of thermal properties on the vapor film thickness is obtained. Based on these findings, we propose a universal law with which we can predict $T_L$ for different liquid drops on substrates with different thermal properties and surface roughness.

\begin{figure}[b]
\centering
\includegraphics[width=0.95\linewidth]{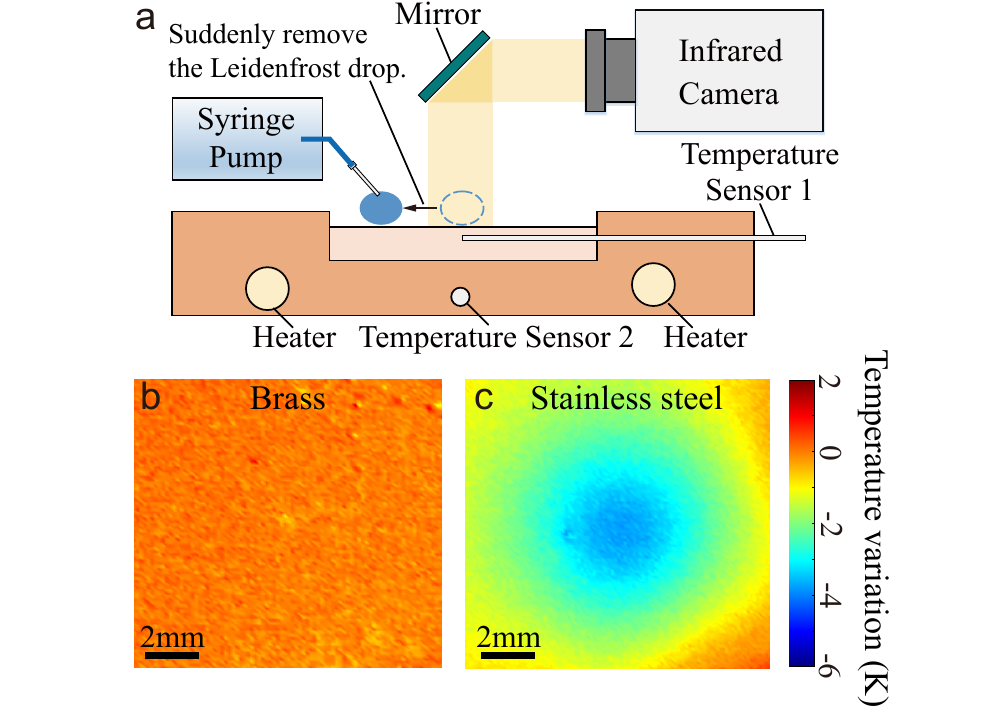}
\caption{The cooling effect of a Leidenfrost drop. \textbf{a.} Schematic of the experimental setup. A water drop is placed on a smooth plate and fixed by a feeding needle for a while. Then the drop is suddenly moved away, and a high-speed infrared camera is used to measure the instantaneous temperature field of the surface. Surface temperature variation of brass (\textbf{b}) and stainless steel (\textbf{c}) surfaces compared to the initial temperature field after the Leidenfrost drop (drop radius $\approx \unit{2}{\milli\meter}$) has evaporated for \unit{60}{\second}.}
\label{fig:1}
\end{figure}

To capture the surface cooling effect caused by a Leidenfrost drop, we place a water drop on a heated substrate with an initial surface temperature $T_{\mathrm{s0}}=\unit{240}{\degreecelsius}$, which is well above the Leidenfrost temperature, and use a 14-gauge metal needle to control the movement of the drop  (see Fig.~\ref{fig:1}a). 
The drop evaporates at a fixed location for $\unit{60}{\second}$, then it is removed with the help of the needle within $\unit{0.05}{\second}$, exposing the surface underneath (see Movie S1).
A high-speed infrared camera (TELOPS FAST L200) immediately records the temperature field of the solid surface. 
The surface temperature recovery during the fast removal process is proved to be less than $\unit{0.8}{\kelvin}$ (see Supplementary Materials for more details).
With the initial homogeneous and known surface temperature, the cooling caused by the drop evaporation can thus be quantitatively characterized.

Fig.~\ref{fig:1}b and Fig.~\ref{fig:1}c compare the temperature reduction of brass and stainless steel surfaces after having a water droplet thereon for 60 seconds. The initial surface temperature is $\unit{296}{\degreecelsius}$ and the drop radius $\approx \unit{2}{\milli\meter}$. 
The stainless steel surface has a maximum temperature reduction of $\unit{6}{\kelvin}$, while the temperature of the brass surface remains almost unchanged (see Fig.~\ref{fig:1}b). The plate temperature reduction due to the evaporation of the levitating Leidenfrost drop was also measured in Ref. \cite {VanLimbeek2017} using an interferometry technique.
Let us focus on the difference in the thermal diffusivity $\alpha$ of these two materials listed in Table \ref{Table1}. $\alpha=k_s/\rho_s c_s$ determines how quickly a material recovers from temperature variations, where $k_s, \rho_s, c_s$ are thermal conductivity, density and specific heat capacity of the solid, respectively. 
It is found that $\alpha$ of stainless steel is about 8 times smaller than brass, which may lead to the difference in surface temperature reduction during the Leidenfrost process.

\begin{table}[t]
	\renewcommand\arraystretch{1.2}
	\setlength{\abovecaptionskip}{0cm}
	\setlength{\belowcaptionskip}{0.3cm}
	\centering  
		\caption{Thermal diffusivity of the selected materials}
 	\setlength{\tabcolsep}{0.5mm}{
	\label{Table1}  
	\begin{tabular}{c c c c c c} 
		\toprule 
		\multicolumn{1}{c}{\multirow{3}{*}{Properties}} & \multicolumn{1}{c}{\multirow{3}{*}{Units}} & \multicolumn{1}{c}{\multirow{2}{*}{Aluminum}} & \multicolumn{1}{c}{\multirow{2}{*}{Brass}} & \multicolumn{1}{c}{1045 Carbon} & \multicolumn{1}{c}{Stainless} \\
\multicolumn{1}{c}{}                            & \multicolumn{1}{c}{}                       & \multicolumn{1}{c}{}                          & \multicolumn{1}{c}{}                       & \multicolumn{1}{c}{Steel}     & \multicolumn{1}{c}{Steel}  \\
\multicolumn{1}{c}{}                            & \multicolumn{1}{c}{}                       & \multicolumn{1}{c}{(Al)}                      & \multicolumn{1}{c}{(Br)}                     & \multicolumn{1}{c}{(CS)}        & \multicolumn{1}{c}{(SS)}     \\

		\midrule 
         Thermal  & \multirow{2}{*}{$\mathrm{m^2/s}$} & 0.907 & 0.356 & 0.136 & 0.045 \\ 
        diffusivity   $ \alpha$ &  & $\mathrm{\times 10^{-4}}$ & $\mathrm{\times 10^{-4}}$ & $\mathrm{\times 10^{-4}}$ & $\mathrm{\times 10^{-4}}$\\
		\bottomrule 
	\end{tabular}}
\end{table}

\begin{figure}[t]
\centering
\includegraphics[width=1\linewidth]{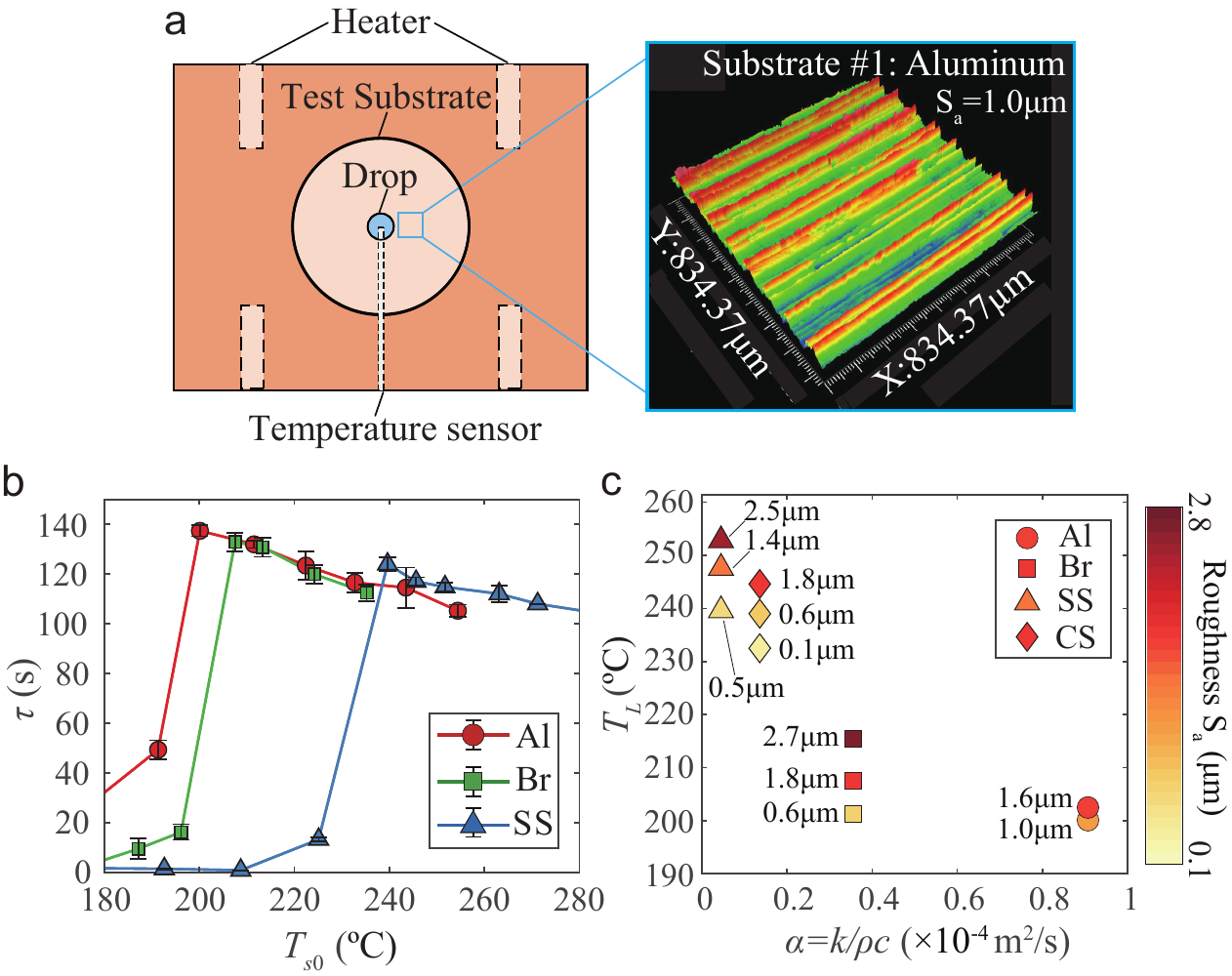}
\caption{ 
Measuring the Leidenfrost temperature $T_L$. 
\textbf{a.} Top view of the experimental setup. A test substrate is placed on the heating base with four heating rods. A temperature sensor is placed at a depth of 2mm from the surface of the test substrate. The inset shows the 3D surface roughness profile of part of a test substrate (Aluminum, $S_\mathrm{a}=\unit{1.0}{\micro\meter}$) obtained by an interference profiler (ZYGONexView).
\textbf{b.} Measured lifetime of a water drop ($V_0=\unit{30.6}{\micro\liter}$) on aluminum (Al), brass (Br) and stainless steel (SS) substrates.
\textbf{c.} The Leidenfrost temperature $T_L$ as a function of thermal diffusivity $\alpha$ and surface roughness ${S_\mathrm{a}}$. The color of the symbols represents the value of ${S_\mathrm{a}}$. The exact values of ${S_\mathrm{a}}$ are labeled beside the symbols.
}
\label{fig:2}
\end{figure}

The demonstrated surface temperature reduction will subsequently affect the vapor generation, and thus the stability of the Leidenfrost stage. To study this dependence, we measure the lifetime $\tau$ of a sessile droplet with a fixed volume $V_0=\unit{30.6}{\micro\liter}$ as a function of the initial surface temperature $T_\mathrm{s0}$ to determine the Leidenfrost temperature \cite{Biance2003}.
The details of the experimental setup are shown in the Supplementary Materials.
Four candidate materials with decreasing thermal diffusivity $\alpha$ - aluminum, brass, carbon steel (Type 1045) and stainless steel (Type 304) - are selected as substrates.
An exemplary surface local surface profile obtained by a 3D interference profiler (ZYGONexView) is shown in Fig.~\ref{fig:2}a.
The surface roughness $S_\mathrm{a}$ is defined as the average surface height deviations from the mean line.
In our experiments, $S_\mathrm{a}$ varies from 0.1 to $\unit{ 2.8}{\micro\meter}$, which is much larger than what typical smooth surfaces have, namely $S_\mathrm{a}\approx\unit{0.02}{\micro\meter}$ \cite{Nagai1996}.

Fig.~\ref{fig:2}b shows the lifetime $\tau$ of a water droplet versus the initial surface temperature $T_\mathrm{s0}$ on substrates of different materials but with a similar roughness ($S_\mathrm{a}= \unit{0.6 \sim 1.0}{\micro\meter}$). 
The Leidenfrost temperature $T_L$ is defined by the specific initial surface temperature  $T_\mathrm{s0}$ that relates to the maximum lifetime $\tau$ of a drop on a superheated surface \cite{Gott1966,Baumeister1973}. 
It is shown that $T_L$ of the droplet on brass is lower than $T_L$ on stainless steel, while higher than $T_L$ on aluminum, which is exactly opposite to the trend of their thermal diffusivity (see Table \ref{Table1}). 
Extending our experiments to surfaces with different roughness, the so-determined Leidenfrost temperatures are plotted as a function of the thermal diffusivity $\alpha$ and the surface roughness $S_\mathrm{a}$ in Fig.~\ref{fig:2}c.
It can be seen, $T_L$ increases with increasing surface roughness for fixed material, and decreases with increasing thermal diffusivity for fixed roughness.

\begin{figure}[t]
\centering
\includegraphics[width=0.9\linewidth]{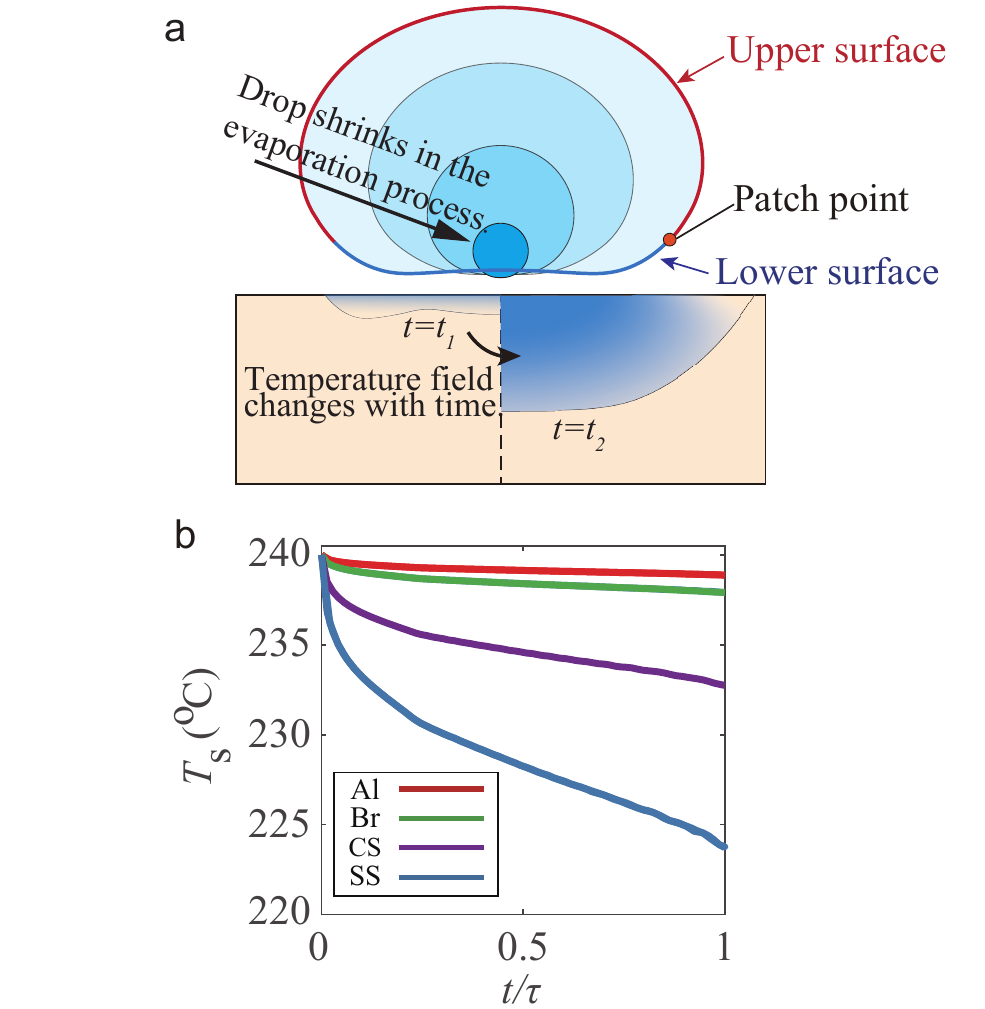}
\caption{
\textbf{a.} The evaporation process of a Leidenfrost drop on a solid substrate. Darker drop colors represent later times.
\textbf{b.} Temporal variation of the surface temperature $T_{\mathrm{s}}$ with initial value $T_{\mathrm{s0}}=\unit{240}{\degreecelsius}$.  
$T_{\mathrm{s}}$ is calculated as the average surface temperature directly below the drop.
}
\label{fig:3}
\end{figure}

For a more quantitative explanation of the experimental observations, we develop a simple theoretical model that simulates the evaporation process of a Leidenfrost droplet from a certain volume till its final fate (see Fig.~\ref{fig:3}a). The temperature variation of the substrate during the evaporation process is considered, while the internal flow and the temperature gradient in the drop are neglected. The evaporation process of the drop is assumed to be quasi-static. 

In the model, the drop geometry is divided into upper and lower regions by the patch point following Ref. \cite{Sobac2014} (see Fig.~\ref{fig:3}a) . While the shape of the upper drop surface is determined by a balance between the hydrostatic pressure and the surface tension, the geometry  of the lower surface is obtained from the thin film equation for the vapor layer within the lubrication approximation, assuming axial symmetry \cite{Sobac2014, VanLimbeek2017, Cai2020a}.
The variation of the temperature profile of the solid substrate $T(r,z)$ during the evaporation process is numerically calculated using the transient heat diffusion equation 
$\frac{1}{\alpha} \frac{\partial T}{\partial t}= \frac{\partial}{\partial r}\left(r \frac{\partial T}{\partial r}\right)+\frac{\partial^{2} T}{\partial z^{2}}$. 
The solid surface is divided into two regions according to the projected area of the lower drop surface. The cooling heat flux $q(r)$ is assumed to only affect the region right beneath the lower droplet surface, while an adiabatic boundary condition is used for the outer region.
A detailed description of this model can be found in the Supplementary Materials (Section 2).

We calculate the evaporation process of a Leidenfrost drop with the initial volume of $V_{\mathrm{0}}=\unit{30.6}{\micro\liter}$ till a very small volume of $V_{\mathrm{end}}=\unit{0.4}{\micro\liter}$, which is close to the size of a Leidenfrost drop approaching its final fate \cite{Celestini2012Take, Lyu2019a}.
Through the calculation, we quantitatively obtain the cooling effect of a Leidenfrost droplet on different surfaces (see Fig.~\ref{fig:3}b).
For materials with high $\alpha$ like aluminum and brass, the surface temperature reduction is less than 2K, while for stainless steel with much smaller $\alpha$ (see Table~\ref{Table1}), the maximum temperature reduction is near $\unit{20}{\kelvin}$.

\begin{figure*}[!t]
\centering
\includegraphics[width=0.85\linewidth]{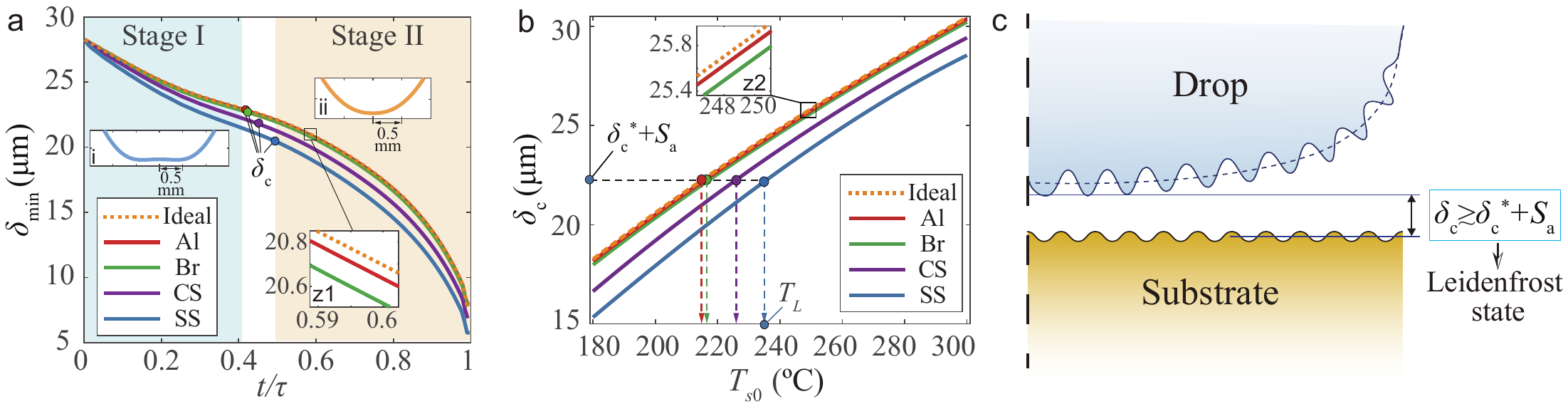}
\caption{
\textbf{a.} Minimum film thickness $\delta_{\mathrm{min}}$ versus normalized time $t/\tau$ on surfaces of different materials with initial temperature $T_{\mathrm{s0}}=\unit{240}{\degreecelsius}$. Inset\,(z1) shows a local zoom of the results to better differentiate the upper three curves. The two stages of the variation of $\delta_{\mathrm{min}}$ are marked using different colors. The bottom geometries of the two stages are shown in inset\,(i) and inset\,(ii), which show the bottom surface with and without a dimple shape, respectively. 
The transition points related to $\delta_{\mathrm{c}}$ of each curve are marked by solid symbols.
\textbf{b.} Characteristic film thickness $\delta_{\mathrm{c}}$ as a function of the initial surface temperature $T_{\mathrm{s0}}$ on aluminum (Al), brass (Br),  carbon steel (CS), stainless steel (SS), and on ideal surfaces. Inset\,(z2) shows a local zoom.
\textbf{c.} A schematic diagram showing the fluctuation profile of the liquid-vapor interface and the solid surface.
}
\label{fig:4}
\end{figure*} 	

Concurrent with the surface cooling, the weaker evaporation flux reduces the vapor pressure in the film, resulting in lower vapor film thickness \cite{VanLimbeek2017}. 
Fig.~\ref{fig:4}a illustrates the variation of the minimum thickness of the vapor film, $\delta_{\mathrm{min}}$, on substrates with $T_{\mathrm{s0}}=\unit{240}{\degree C}$ of various materials, including an ideal surface ($\alpha\to \infty$). 
It can be seen that with decreasing $\alpha$ for realistic materials, the minimum film thickness $\delta_{\mathrm{min}}$ deviates further from the ideal result. 
The difference is quite small for aluminum and brass surfaces due to their relatively high thermal diffusivities, while much larger for carbon steel and stainless steel surfaces. This trend holds for all initial surface temperatures $T_{\mathrm{s0}}$.

The temporal variation of $\delta_{\mathrm{min}}$ in Fig.~\ref{fig:4}a can be divided into two stages according to the lower surface geometry of the drop.
With the evaporation process, the bottom surface loses its dimple shape and the minimum position of the film thickness moves to the center, indicating the transition from the first stage to the second stage.
The minimum film thickness at the transition moment is defined as the characteristic film thickness $\delta_{\mathrm{c}}$.
In the second stage, the spatial and temporal fluctuations of the liquid-vapor interface are suppressed because now the surface tension takes the dominance in defining the geometry \cite{Graeber2021}, resulting in more stable levitation of the Leidenfrost drop until its final fate.
Therefore, $\delta_{\mathrm{c}}$ represents the minimum film thickness before the stable period in the Leidenfrost drop evaporation process. Fig.~\ref{fig:4}b shows that $\delta_{\mathrm{c}}$ increases with initial temperature $T_{\mathrm{s0}}$ and thermal diffusivity $\alpha$ of the substrate.

As an inherently unstable system, the Leidenfrost drop is known to suffer from various instabilities \cite{Chantelot2021, Taylor1950, Panzarella2000,2007Liquid, Aursand2018, Zhao2020,  Chantelot2021Drop}, which induce fluctuations with an amplitude in the order of $\unit{5}{\micro\meter}$ at the liquid-vapor interface \cite{Graeber2021}.
The fluctuation, combined with the roughness of the solid surface, results in significant inhomogeneity of the local vapor thickness $\delta$  (see Fig.~\ref{fig:4}c).
Under the influence of the wavy geometry of both the drop and the solid surface, the associated overpressure of the vapor layer $\Delta P$ experiences strong variation \cite{Graeber2021}, which causes further fluctuation of the liquid-gas interface.
As the characteristic film thickness $\delta_{\mathrm{c}}$ becomes smaller, the undulated liquid interface is more likely to penetrate the vapor film and directly contact with the heated surface \cite{Biance2003, Chantelot2021}.
Direct contact leads to strong nucleate boiling of the drop and further increase of the contact area \cite{Chantelot2021,Harvey2021, Chantelot2021Drop}, bringing the Leidenfrost stage to an end \cite{Tran2012PRL}.

To prevent such collapse of the vapor film and to maintain the Leidenfrost state, the characteristic film thickness $\delta_{\mathrm{c}}$ must be larger than a critical value. 
For a perfectly smooth surface, this critical value is defined as $\delta_{\mathrm{c}}^*$, which relates only  to the liquid properties. 
For surfaces with finite roughness, the film thickness must be increased to compensate for the effect of roughness. Here we propose the criterion $\delta_{\mathrm{c}} \gtrsim \delta_{\mathrm{c}}^* + S_\mathrm{a}$, which considers a simple linear effect of the roughness on the critical film thickness required to maintain the Leidenfrost stage, and is consistent with the view of the Leidenfrost transition as a directed percolation process \cite{Chantelot2021}.

Using the criterion $\delta_{\mathrm{c}} \gtrsim \delta_{\mathrm{c}}^* + S_\mathrm{a}$, we can explain how thermal diffusivity and surface roughness affect the Leidenfrost temperature $T_L$.
As the surface roughness $S_\mathrm{a}$ increases, a thicker vapor film is required to maintain the Leidenfrost state.  Since the characteristic film thickness $\delta_{\mathrm{c}}$ monotonically increases with the initial surface temperature $T_{\mathrm{s0}}$, the increment of $S_\mathrm{a}$ leads to a higher $T_L$, as evidenced by the above experiments.
The effects of the thermal diffusivity $\alpha$ on $T_L$ can be understood through $\delta_{\mathrm{c}}(T_\mathrm{s0})$  dependencies that refer to different thermal diffusivities in Fig.~\ref{fig:4}b.
For poorly diffusive materials such as stainless steel, the decline in surface temperature reduces the characteristic vapor layer thickness $\delta_\mathrm{c}$, and a higher initial surface temperature is required for sufficient film thickness to meet the stability criterion.
As a result, the $T_L$ decreases with increasing thermal diffusivity $\alpha$. However, for materials with relatively high $\alpha$ like aluminum and brass, the cooling effect is negligible and further increases in $\alpha$ have little effect on $T_L$.

\begin{figure}[h]
\flushleft
\centerline{\includegraphics[width=0.9\linewidth]{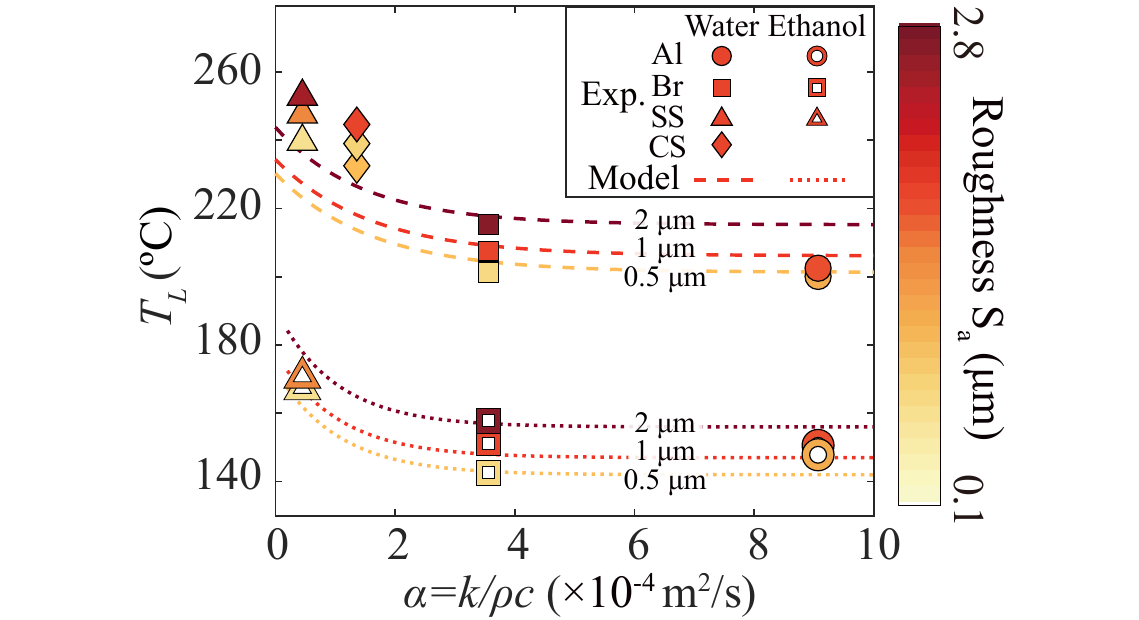}}
\caption{
$T_L$ resulting from our model for water (dashed line) and ethanol drops (dotted line) on surfaces with different roughness ($S_\mathrm{a}=\unit{2,1,0.5}{\micro\meter}$), compared to experimentally measured values of water (solid markers) and ethanol drops (hollow markers). The colors of the markers correspond to the value of the surface roughness $S_\mathrm{a}$, which is also given for the curves.
}
\label{fig:5}
\end{figure}

To obtain the $T_L$ for surfaces of different materials and surface roughness, we interpolate the value of $\delta_{\mathrm{c}}^* + S_\mathrm{a}$ into the $\delta_{\mathrm{c}}(T_\mathrm{s0})$  dependencies to get the transition temperature (see the illustration in Fig.~\ref{fig:4}b).
Here we choose three typical values of surface roughness ($S_\mathrm{a}=\unit{0.5, 1, 2}{\micro\meter}$) to show a basic trend of the effect of roughness.
The exact value of $\delta_{\mathrm{c}}^*$ can be approximated by experimentally measuring the Leidenfrost temperature $T_L$ on an extremely low roughness surface and interpolating the obtained $T_L$ into the $\delta_{\mathrm{c}}(T_\mathrm{s0})$ dependency.

The obtained $T_L$ values are fitted with the function $y=a e^{bx+c}+d$ and  plotted in Fig.~\ref{fig:5} as dashed lines, together with the experimental results. 
The experimental and model results are also shown for ethanol droplets to illustrate the generality of our model. 
As can be seen, our simple model shows good agreement with the experimental data.
Moreover, the model predicts that $T_L$ decreases with increasing $\alpha$ for the same liquid and gradually converges to a nearly constant temperature, as we have already speculated.

In summary, we showed by infrared thermometry the non-negligible cooling effects of Leidenfrost drops on surfaces with small thermal diffusivities. 
The influence of the thermal properties and surface roughness on the Leidenfrost temperature was then investigated experimentally and explained by a simple theoretical model.
We propose the criterion $\delta_{\mathrm{c}} \gtrsim \delta_{\mathrm{c}}^* + S_\mathrm{a}$ based on the stability of the vapor film in Leidenfrost stage. Combining the model and the experimental results, we showed that this model can be extended to different fluids and solid substrates.
In future work, the present model should be extended to the dynamic Leidenfrost effect \cite{Tran2012PRL} by considering the impact process of the droplet to help to predict and control the Leidenfrost temperature in practical applications.

 \begin{acknowledgments}
This work was supported by National Natural Science Foundation of China under Grants Nos. 11988102, 91852202, 51976105, 91841302, Tencent Foundation through the XPLORER PRIZE, ERC Advanced Grant DDD, Project No. 740479, and NWO through the Multiscale Catalytic Energy Conversion (MCEC) Research Center.
 \end{acknowledgments}

\subsection{Data availability.} The data that support the findings of this study are available from the corresponding author upon reasonable request.
~

%\bibliographystyle{prsty_withtitle_withoutline} 
%\bibliography{Literature.bib}

\begin{thebibliography}{10}

\bibitem{Biance2003}
A.-L. Biance, C. Clanet, and D. Qu{\'e}r{\'e}, {\it Leidenfrost drops}, Phys.
  Fluids {\bf 15},  1632  (2003).

\bibitem{Quere2013}
D. Qu{\'e}r{\'e}, {\it Leidenfrost dynamics}, Annu. Rev. Fluid Mech. {\bf 45},
  197  (2013).

\bibitem{Chantelot2021}
P. Chantelot and D. Lohse, {\it Leidenfrost effect as a directed percolation
  phase transition}, Phys. Rev. Lett. {\bf 127},  124502  (2021).

\bibitem{Bouillant2021Self}
A. Bouillant, C. Cohen, C. Clanet, and D. Qu{\'e}r{\'e}, {\it Self-excitation
  of Leidenfrost drops and consequences on their stability}, Proc. Natl. Acad.
  Sci. U. S. A. {\bf 118},  e2021691118  (2021).

\bibitem{Sijia2021On}
S. Lyu, H. Tan, Y. Wakata, X. Yang, C.~K. Law, D. Lohse, and C. Sun, {\it On
  explosive boiling of a multicomponent Leidenfrost drop}, Proc. Natl. Acad.
  Sci. U. S. A. {\bf 118},  e2016107118  (2021).

\bibitem{Kim2007}
J. Kim, {\it Spray cooling heat transfer: The state of the art}, Int. J. Heat
  Fluid Flow {\bf 28},  753  (2007).

\bibitem{Liang2017}
G. Liang and I. Mudawar, {\it Review of spray cooling--Part 2: High temperature
  boiling regimes and quenching applications}, Int. J. Heat Mass Transf. {\bf
  115},  1206  (2017).

\bibitem{Vakarelski2016}
I.~U. Vakarelski, J.~D. Berry, D.~Y. Chan, and S.~T. Thoroddsen, {\it
  Leidenfrost vapor layers reduce drag without the crisis in high viscosity
  liquids}, Phys. Rev. Lett. {\bf 117},  114503  (2016).

\bibitem{Gauthier2019}
A. Gauthier, C. Diddens, R. Proville, D. Lohse, and D. van~der Meer, {\it
  Self-propulsion of inverse Leidenfrost drops on a cryogenic bath}, Proc.
  Natl. Acad. Sci. U. S. A. {\bf 116},  1174  (2019).

\bibitem{Bouillant2018}
A. Bouillant, T. Mouterde, P. Bourrianne, A. Lagarde, C. Clanet, and D.
  Qu{\'e}r{\'e}, {\it Leidenfrost wheels}, Nat. Phys. {\bf 14},  1188  (2018).

\bibitem{Abdelaziz2013}
R. Abdelaziz, D. Disci-Zayed, M.~K. Hedayati, J.-H. P{\"o}hls, A.~U. Zillohu,
  B. Erkartal, V.~S.~K. Chakravadhanula, V. Duppel, L. Kienle, and M. Elbahri,
  {\it Green chemistry and nanofabrication in a levitated Leidenfrost drop},
  Nat. Commun. {\bf 4},  1  (2013).

\bibitem{Nagai1996}
N. Nagai and S. Nishio, {\it Leidenfrost temperature on an extremely smooth
  surface}, Exp. Therm. Fluid Sci. {\bf 12},  373  (1996).

\bibitem{Kim2011}
H. Kim, B. Truong, J. Buongiorno, and L.-W. Hu, {\it On the effect of surface
  roughness height, wettability, and nanoporosity on Leidenfrost phenomena},
  Appl. Phys. Lett. {\bf 98},  083121  (2011).

\bibitem{Jiang2022Inhibiting}
M. Jiang, Y. Wang, F. Liu, H. Du, Y. Li, H. Zhang, S. To, S. Wang, C. Pan, J.
  Yu, {\it et~al.}, {\it Inhibiting the Leidenfrost effect above 1,000° C for
  sustained thermal cooling}, Nature {\bf 601},  568  (2022).

\bibitem{Wu2021}
R. Wu, O. Lamini, and C. Zhao, {\it Leidenfrost temperature: surface thermal
  diffusivity and effusivity effect}, Int. J. Heat Mass Transf. {\bf 168},
  120892  (2021).

\bibitem{Celestini2013}
F. Celestini, T. Frisch, and Y. Pomeau, {\it Room temperature water Leidenfrost
  droplets}, Soft Matter {\bf 9},  9535  (2013).

\bibitem{VanLimbeek2021}
M.~A. van Limbeek, O. Ram{\'\i}rez-Soto, A. Prosperetti, and D. Lohse, {\it How
  ambient conditions affect the Leidenfrost temperature}, Soft Matter {\bf 17},
   3207  (2021).

\bibitem{Baumeister1973}
K.~J. Baumeister and F.~F. Simon, {\it {Leidenfrost Temperature—Its
  Correlation for Liquid Metals, Cryogens, Hydrocarbons, and Water}}, J. Heat
  Transf. {\bf 95},  166  (1973).

\bibitem{Bourrianne2019}
P. Bourrianne, C. Lv, and D. Qu{\'e}r{\'e}, {\it The cold Leidenfrost regime},
  Sci. Adv. {\bf 5},  eaaw0304  (2019).

\bibitem{Panchanathan2021}
D. Panchanathan, P. Bourrianne, P. Nicollier, A. Chottratanapituk, K.~K.
  Varanasi, and G.~H. McKinley, {\it Levitation of fizzy drops}, Sci. Adv. {\bf
  7},  eabf0888  (2021).

\bibitem{Bernardin1999}
J.~D. Bernardin and I. Mudawar, {\it {The Leidenfrost Point: Experimental Study
  and Assessment of Existing Models}}, J. Heat Transf. {\bf 121},  894  (1999).

\bibitem{Panzarella2000}
C.~H. Panzarella, S.~H. Davis, and S.~G. Bankoff, {\it Nonlinear Dynamics in
  Horizontal Film Boiling}, J. Fluid Mech. {\bf 402},  163  (2000).

\bibitem{Aursand2018}
E. Aursand, S.~H. Davis, and T. Ytrehus, {\it Thermocapillary Instability as a
  Mechanism for Film Boiling Collapse}, J. Fluid Mech. {\bf 852},  283  (2018).

\bibitem{Zhao2020}
T.~Y. Zhao and N.~A. Patankar, {\it The thermo-wetting instability driving
  Leidenfrost film collapse}, Proc. Natl. Acad. Sci. U. S. A. {\bf 117},  13321
   (2020).

\bibitem{Cai2020a}
C. Cai, I. Mudawar, H. Liu, and C. Si, {\it Theoretical Leidenfrost point (LFP)
  model for sessile droplet}, Int. J. Heat Mass Transf. {\bf 146},  118802
  (2020).

\bibitem{VanLimbeek2017}
M.~A. Van~Limbeek, M.~H.~K. Schaarsberg, B. Sobac, A. Rednikov, C. Sun, P.
  Colinet, and D. Lohse, {\it Leidenfrost drops cooling surfaces: theory and
  interferometric measurement}, J. Fluid Mech. {\bf 827},  614  (2017).

\bibitem{Gott1966}
B. Gottfried, C. Lee, and K. Bell, {\it The Leidenfrost phenomenon: film
  boiling of liquid droplets on a flat plate}, Int. J. Heat Mass Transf. {\bf
  9},  1167  (1966).

\bibitem{Sobac2014}
B. Sobac, A. Rednikov, S. Dorbolo, and P. Colinet, {\it Leidenfrost effect:
  Accurate drop shape modeling and refined scaling laws}, Phys. Rev. E {\bf
  90},  053011  (2014).

\bibitem{Celestini2012Take}
F. Celestini, T. Frisch, and Y. Pomeau, {\it Take off of small Leidenfrost
  droplets}, Phys. Rev. Lett. {\bf 109},  034501  (2012).

\bibitem{Lyu2019a}
S. Lyu, V. Mathai, Y. Wang, B. Sobac, P. Colinet, D. Lohse, and C. Sun, {\it
  Final fate of a Leidenfrost droplet: Explosion or takeoff}, Sci. Adv. {\bf
  5},  eaav8081  (2019).

\bibitem{Graeber2021}
G. Graeber, K. Regulagadda, P. Hodel, C. K{\"u}ttel, D. Landolf, T.~M.
  Schutzius, and D. Poulikakos, {\it Leidenfrost droplet trampolining}, Nat.
  Commun. {\bf 12},  1  (2021).

\bibitem{Taylor1950}
G.~I. Taylor, {\it The instability of liquid surfaces when accelerated in a
  direction perpendicular to their planes. I}, Proc. R. Soc. London, Ser. A
  {\bf 201},  192  (1950).

\bibitem{2007Liquid}
V.~P. Carey, {\it Liquid-vapor phase-change phenomena: an introduction to the
  thermophysics of vaporization and condensation processes in heat transfer
  equipment} (CRC Press, Boca Raton, 3rd Edition, 2020).

\bibitem{Harvey2021}
D. Harvey, J.~M. Harper, and J.~C. Burton, {\it Minimum Leidenfrost temperature
  on smooth surfaces}, Phys. Rev. Lett. {\bf 127},  104501  (2021).

\bibitem{Chantelot2021Drop}
P. Chantelot and D. Lohse, {\it Drop impact on superheated surfaces: short-time
  dynamics and transition to contact}, J. Fluid Mech. {\bf 928},  A36  (2021).

\bibitem{Tran2012PRL}
T. Tran, H.~J.~J. Staat, A. Prosperetti, C. Sun, and D. Lohse, {\it Drop Impact
  on Superheated Surfaces}, Phys. Rev. Lett. {\bf 108},  036101  (2012).

\end{thebibliography}

\end{document}